\documentclass[12pt]{article}

\usepackage[english]{babel}
\usepackage[utf8]{inputenc}
\usepackage{amsmath}
\usepackage{graphicx,geometry}
\usepackage{subfig}
\usepackage[noblocks]{authblk}
\usepackage{hyperref}
\newcommand{\be}{\begin{equation}}
\newcommand{\ee}{\end{equation}}

\title{Chiral phase transition at finite chemical potential in 2+1-flavor soft-wall AdS/QCD}
\author[1]{Sean P. Bartz}
\author[2]{Theodore Jacobson}
\affil[ ]{\emph{Dept.~of Physics and Astronomy, Macalester College, St.~Paul, MN 55105}}
\affil[1]{\href{mailto:sbartz@macalester.edu}{sbartz@macalester.edu} }
\affil[2]{\href{mailto:tjacobs1@macalester.edu}{tjacobs1@macalester.edu} }

\date{\today}
\begin{document}
\maketitle

\begin{abstract}

The phase transition from hadronic matter to chirally-symmetric quark-gluon plasma is expected to be a rapid crossover at zero quark chemical potential ($\mu$), becoming  first order at some finite value of $\mu$,  indicating the presence of a critical point. 
Using a three-flavor soft-wall model of AdS/QCD, we investigate the effect of varying the light and strange quark masses on the order of the chiral phase transition. 
%In this modified soft-wall model, higher-order terms in the scalar potential allow for independent sources of explicit and spontaneous symmetry breaking and admit first order phase transitions at a critical temperature $T_c$ which depends on quark chemical potential. 
%We analyze the $SU(3)$ case where $m_{u/d}=m_s$ as well as the non-degenerate case $m_{u/d}\not=m_s$. 
At zero quark chemical potential, we reproduce the Columbia Plot, which summarizes the results of lattice QCD and other holographic models.
We then extend this holographic model to examine the effects of finite quark  chemical potential.
We find that the the chemical potential does not affect the critical line that separates first-order from rapid crossover transitions. 
This excludes the possibility of a critical point in this model, suggesting that a different setup is necessary to reproduce all the features of the QCD phase diagram. 
\end{abstract}

\clearpage

%============================================

%============================================

\section{Introduction}
The investigation of the phase diagram for quark matter is a major project of nuclear physics. In particular,  heavy-ion community is interested in mapping the phase transition between hadronic matter and the quark-gluon plasma as a function of temperature and baryon chemical potential. In particular, the search for the expected critical point in this phase boundary is the focus of the next run at the Relativistic Heavy Ion Collider (RHIC) at Brookhaven National Laboratory. A robust theoretical description of the phase boundary is thus a worthy goal.

Results from lattice QCD have demonstrated that the deconfinement phase transition at zero quark chemical potential ($\mu$) is a rapid crossover. Extending lattice results to finite chemical potential is prevented by a well-known obstacle known as the sign problem. Techniques exist for extending lattice QCD to finite but relatively small values of $\mu$, results which show no evidence of a critical point but place limits upon its location \cite{deForcrand:2006pv,Bonati2014,Bellwied2016}. 

Phenomenological models inspired by the AdS/CFT correspondence \cite{Maldacena1998TheSupergravity, Witten1998Anti-deTheories, karch-katz-son-adsqcd}  have succeeded in describing some aspects of the quark-gluon plasma \cite{Vega2017, Zollner2017}. Previous work has mapped the chiral phase transition as a function of temperature and quark chemical potential \cite{Bartz:2016ufc, Cui2013ThermalPotential, Colangelo2012TemperatureStudy,Chelabi2016ChiralAdS/QCD,Fang2016ChiralStudy,Fang2016ChiralAdS/QCD}. In two-flavor models, the phase transition is a crossover for zero quark mass and second order for nonzero quark mass. This is consistent with lattice QCD models, which find the order of the order of the chiral phase transition to be dependent on the light and strange quark masses. 
%This is summarized in the Columbia plot, which indicates that a first-order transition is achievable in 2+1-flavor models.
Some holographic models have shown evidence of a critical point by examining baryon susceptibilities\cite{Critelli2017,Critelli2017a,Knaute2017,Li2018}, but these models do not examine chiral symmetry restoration, which is the focus of this work. 

\begin{figure}[htb] 
\centering
  \includegraphics[width=0.8\textwidth]{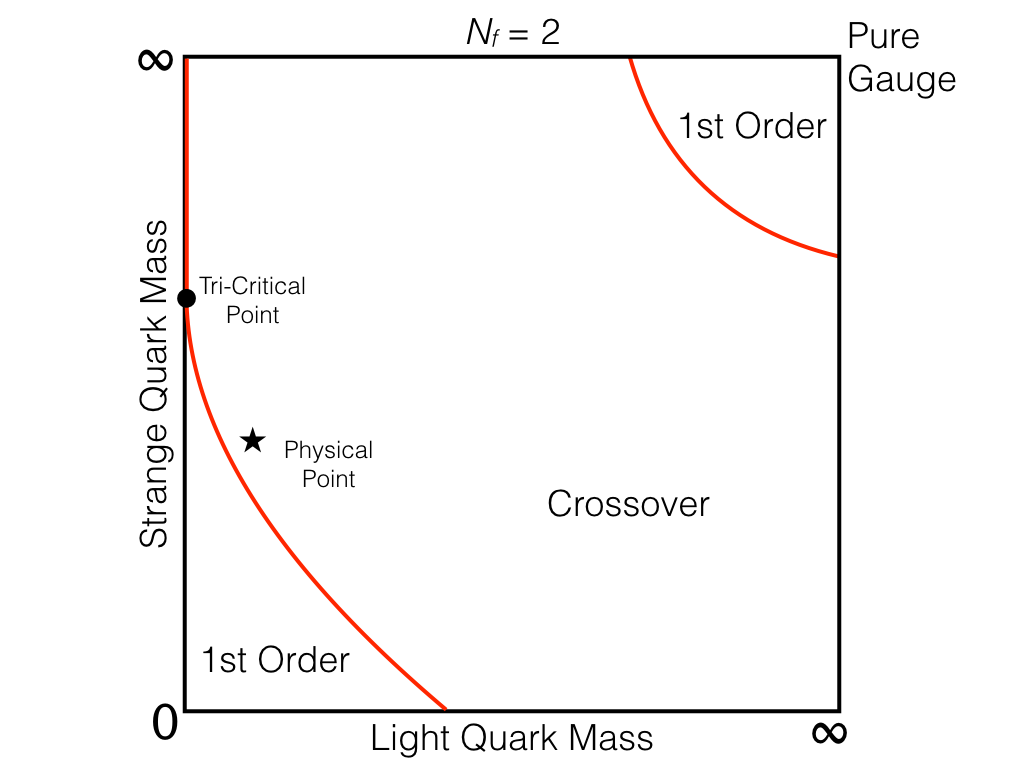}
\caption{A sketch of the expected Columbia Plot, indicating the order of the chiral phase transition as a function of light and strange quark masses at zero chemical potential. The crossover region is separated from regions of first-order phase transitions by second-order phase transition lines.}
\label{columbia}
\end{figure}

\begin{figure}[htb] 
\centering
  \includegraphics[width=0.8\textwidth]{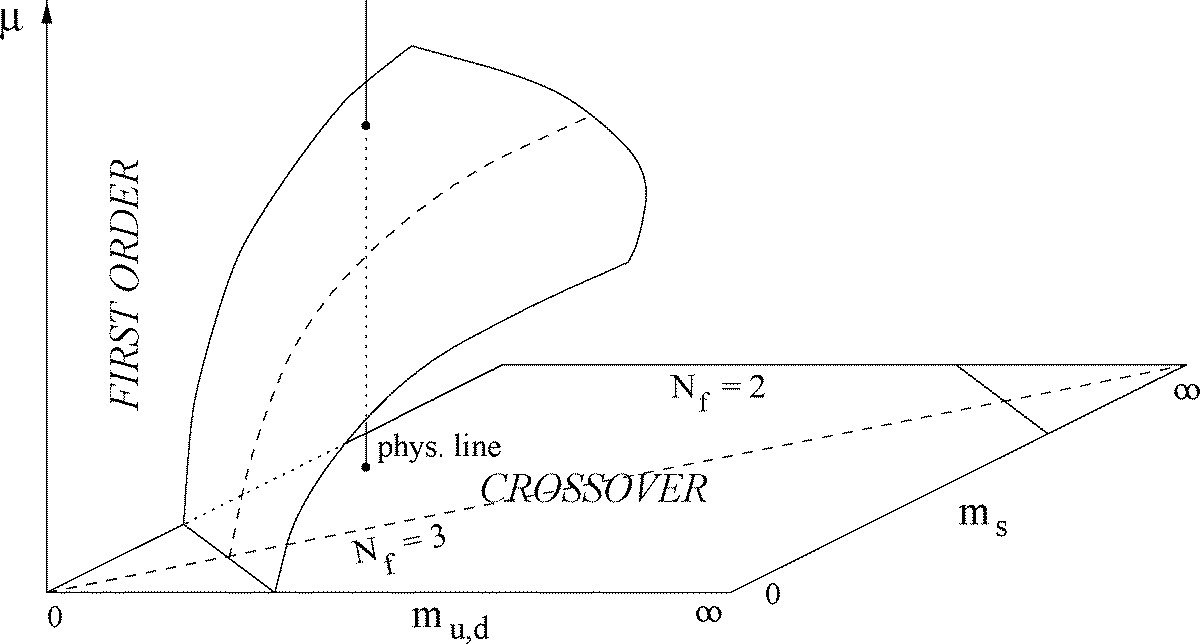}
\caption{The critical surface separates first order from crossover phase transitions. Image taken from \cite{Laermann:2003cv}.}
\label{critical_surface}
\end{figure}

The chiral phase structure for varying quark masses is summarized by the Columbia Plot \cite{Brown1990,Laermann:2003cv}, sketched in Figure \ref{columbia}.   Visualizing $\mu$ as a third axis of the Columbia Plot, the 2nd-order line becomes a critical surface. 
To achieve the expected critical point in the $T-\mu$ plane, the critical surface should have the curvature shown in Figure \ref{critical_surface}, with the physical point being in the crossover region at $\mu=0$.The behavior of a pure gauge theory, found in the upper-right corner of the plot, is beyond the scope of this work.  

Our previous work \cite{Bartz:2016ufc} examined light quarks only, corresponding to the $N_f=2$ line of the Columbia Plot, where $m_s$ is effectively infinite. In this regime, the chiral phase transition is a rapid crossover for  finite quark mass and second-order for massless quarks.
The authors of \cite{Li2017} studied the Columbia Plot at zero chemical potential in a soft-wall model of AdS/QCD.
In this paper, we reproduce these results, and extend this holographic analysis to finite chemical potential.

% This paper is organized as follows.
% In section \ref{model} we review the setup of the soft-wall AdS/QCD model and the background metric, showing how these are modified when considering 2+1 quark flavors.
% In Section \ref{chiral} we  analyze the chiral phase transition in the $SU(3)$ case where $m_l=m_s$ and the  2+1-flavor case where $m_l \neq m_s$ at finite quark chemical potential. 
% %In particular we are interested in the existence of a critical surface which separates first-order from crossover transitions, as depicted in Figure \ref{critical_surface}. 
% Notably, the quark chemical potential is coupled to both light and heavy quarks, as determined by the background metric.  
% Finally, in Section \ref{chipotential} we analyze the relationship of the chiral potential to the type of phase transitions produced.

\section{Soft-Wall Model}\label{model}
To consider the thermodynamics of AdS/QCD, we use an asymptotically anti-de Sitter  5-D black hole metric 
\be
ds^2=\frac{L^2}{z^2}\left(-f(z)dt^2+dx_i^2+\frac{dz^2}{f(z)}\right).\label{metricGeneric}
\ee
Following the procedure established in our previous work and in \cite{Chamblin1999ChargedHolography,Park2010DissociationMedium,Colangelo2011HolographyDiagram},  we model finite temperature and chemical potential with a charged black hole described by the  5D AdS--Reissner-Nordstr{\"o}m metric
\be 
f(z)=1-(1+Q^2)\left(\frac{z}{z_h}\right)^4+Q^2\left(\frac{z}{z_h}\right)^6, \label{metricMu}
\ee
where $z_h$ is th location of the event horizon and  $Q=qz_h^3$ is the black hole charge with the constraint $0<Q^2<2$. The chemical potential and temperature are uniquely determined by the charge and horizon position
\begin{eqnarray}
\mu&=&\kappa \frac{Q}{z_h}, \\
T&=&\frac{1}{\pi z_h}\left(1-\frac{Q^2}{2}\right).
\end{eqnarray} 
Note that $\mu$ is the quark chemical potential, with a value one third of the baryon chemical potential. As in \cite{Colangelo2012TemperatureStudy} we take $\kappa=1$. %Any combination of temperature and chemical potential uniquely determines a combination of $z_h$ and $q$. 

The matter fields are described by the action
\be
\mathcal{S}=\frac{1}{2k} \int d^5x \sqrt{-g} e^{-\Phi(z)}\  \mathrm{Tr}\left[|DX|^2+V_m(X)\right],\label{fullaction}
\ee
where $X$ contains the scalar and pseudoscalar meson fields. We exclude the vector and axial-vector meson fields to focus on chiral dynamics. The vacuum expectation value (VEV) of the scalar field describes the chiral symmetry breaking of the model. In a 2-flavor symmetric model, the VEV is given by
\be
\langle X \rangle = \frac{\chi(z)}{2}I,
\ee
where $I$ is the $N_f \times N_f$ identity matrix. 
Allowing for flavor asymmetry, the scalar VEV takes the form
\begin{equation}
\langle X \rangle = \begin{pmatrix}
\frac{\chi_l(z)}{\sqrt{2}} & 0 & 0 \\
0 & \frac{\chi_l(z)}{\sqrt{2}} & 0 \\
0 & 0 & \frac{\chi_s(z)}{\sqrt{2}}
\end{pmatrix},
\end{equation}
where the normalization factor is chosen to give the kinetic term its canonical form.

The features of this particular model are established by the choice of scalar potential $V_m(X)$. 
The AdS/CFT dictionary establishes the mass term, and a quartic term is necessary to obtain independent sources of explicit and spontaneous chiral symmetry breaking \cite{karch-katz-son-adsqcd,gherghetta-kelley,Bartz2014DynamicalModel}.  We also include t'Hooft determinant term in the scalar potential to introduce flavor mixing \cite{Chelabi2016ChiralAdS/QCD}. The potential becomes
\begin{equation}
V_m(X) = m_5^2 |X|^2 + 4v_4 |X|^4 + \gamma \textrm{Re}\left[\textrm{det}(X) \right],
\end{equation}
where $\gamma = 6 \sqrt{2}\ v_3$. As before, we take $v_4=8$, and in the following set $v_3 = -3$.

\section{Chiral Symmetry restoration}\label{chiral}

To examine the chiral dynamics, we consider the behavior of the background chiral fields. The chiral potential is found to be
\begin{equation}
V(\chi) = \langle \textrm{Tr}[ V_m(X) ] \rangle = m_5^2 \left(\chi_l^2 + \frac{1}{2}\chi_s^2 \right) + 3v_3 \chi_l^2\chi_s + v_4 (2\chi_l^4+\chi_s^4). \label{chipotential}
\end{equation}
Varying the action (\ref{fullaction})  yields the equations of motion
\begin{align}
\chi_l'' - \left(\frac{3f(u)-uf'(u) + uf(u)\Phi'(u)}{uf(u)} \right) \chi_l' + \frac{1}{u^2f(u)} \left(3\chi_l - 3 v_3 \chi_l\chi_s - 4v_4 \chi_l^3 \right) &=0, \label{2+1EOM1} \\
\chi_s'' - \left(\frac{3f(u)-uf'(u) + uf(u)\Phi'(u)}{uf(u)} \right) \chi_s' + \frac{1}{u^2f(u)} \left(3\chi_s - 3 v_3 \chi_l^2 - 4v_4 \chi_s^3 \right) &=0,
\label{2+1EOM2}
\end{align}
where we have changed variables to $u=z/z_h$. The coupling of the light and strange sectors is evident, and vanishes in the case $v_3=0$. 
The UV boundary conditions on the chiral fields are determined by the AdS/CFT dictionary
\begin{equation}
\chi_l(u \rightarrow 0) = m_l \zeta z_h u + \frac{\sigma_l}{\zeta} z_h^3u^3, \ \ \chi_s(u \rightarrow 0) = m_s \zeta z_h u + \frac{\sigma_s}{\zeta} z_h^3u^3, \label{chiralUV}
\end{equation}
where  $m_l = m_{u,d}$ is the light quark mass, $m_s$ is the strange quark mass, and there are two chiral condensates  $\sigma_l = \langle \bar{u} u \rangle = \langle \bar{d} d \rangle$ and $\sigma_s = \langle \bar{s}s \rangle$. The normalization $\zeta=\sqrt{N_c}/(2\pi)$ is determined from large-$N$ QCD \cite{Cherman:2008eh}. 
The quark masses are taken as input parameters, with the chiral condensates determined by numerically solving the boundary value problem (\ref{2+1EOM1},\ref{2+1EOM2}) with the near-horizon condition that the chiral fields remain finite.
For a given quark mass, we solve for the chiral condensates as a function of temperature and baryon density. Previous work shows that the $T$ and $\mu$ dependence of the phase transition is qualitatively equivalent \cite{Bartz:2016ufc}. For clarity of illustration, we will plot $\sigma_l, \, \sigma_s$ as functions of temperature only.

\subsection{Flavor-symmetric case\label{3flav}}

We begin our analysis with the flavor-symmetric case. This is represented by the dashed line $m_l = m_s$ on the Columbia Plot. In this case, $\chi_l = \chi_s$, reducing (\ref{2+1EOM1},\ref{2+1EOM2}) to the single equation of motion
\begin{equation}
\chi'' - \left(\frac{3f(u)-uf'(u)+ uf(u)\Phi'(u)}{uf(u)} \right) \chi' + \frac{1}{u^2f(u)} \left(3\chi - 3 v_3 \chi^2 - 4v_4 \chi^3 \right) =0, \label{symmetricEOM}
\end{equation}
subject to the UV boundary condition \ref{chiralUV}. Using a numerical shooting method, we find the value of $\sigma$ that produces solutions to (\ref{symmetricEOM}) that are regular at the black hole horizon. When nontrivial solutions exist, they are energetically favored over the trivial solution $\chi(u)=0$ \cite{Chelabi2016ChiralAdS/QCD}. 
Chiral symmetry restoration is realized at values of $T,\, \mu$ where nontrivial regular solutions  do not exist, making the trivial solution the only solution and signifying $\sigma =0$.

Although this version of the model is flavor-symmetric, it does not simply mimic the two-flavor results of \cite{Bartz:2016ufc}, which represent the top line of the Columbia Plot, where $m_s$ is effectively infinite. The inclusion of the cubic t'Hooft determinant term in the scalar potential breaks the $\chi \leftrightarrow -\chi$ symmetry, giving energetic preference to positive chiral fields \cite{Chelabi2016ChiralAdS/QCD}. As a result, the $\sigma \leftrightarrow -\sigma$ symmetry found in the two-flavor case is also broken, allowing first-order phase transitions for some values of the quark mass. 

\begin{figure}[htb] 
\centering
\subfloat[]{
  \includegraphics[width=0.33\textwidth]{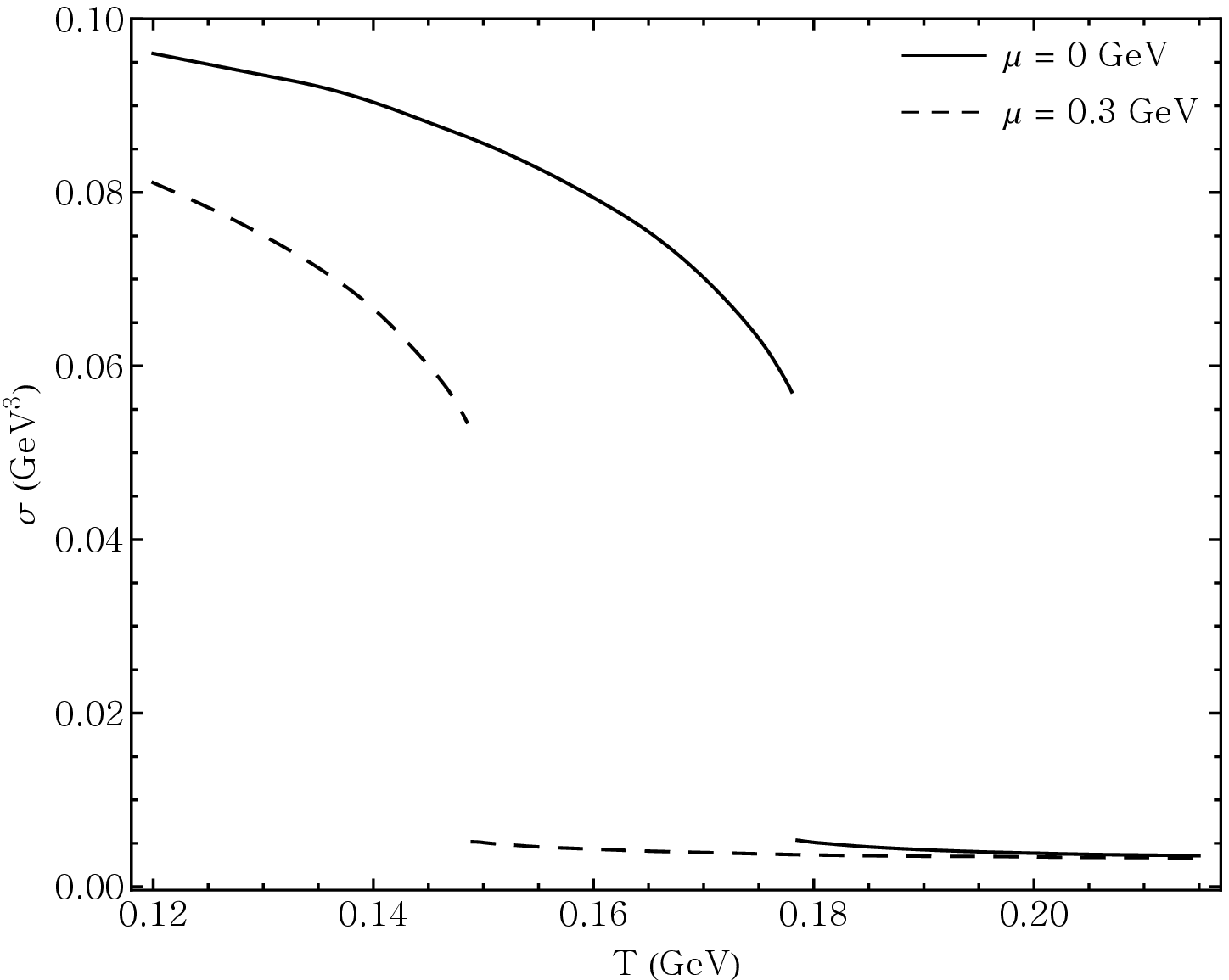}
}
\subfloat[]{
  \includegraphics[width=0.33\textwidth]{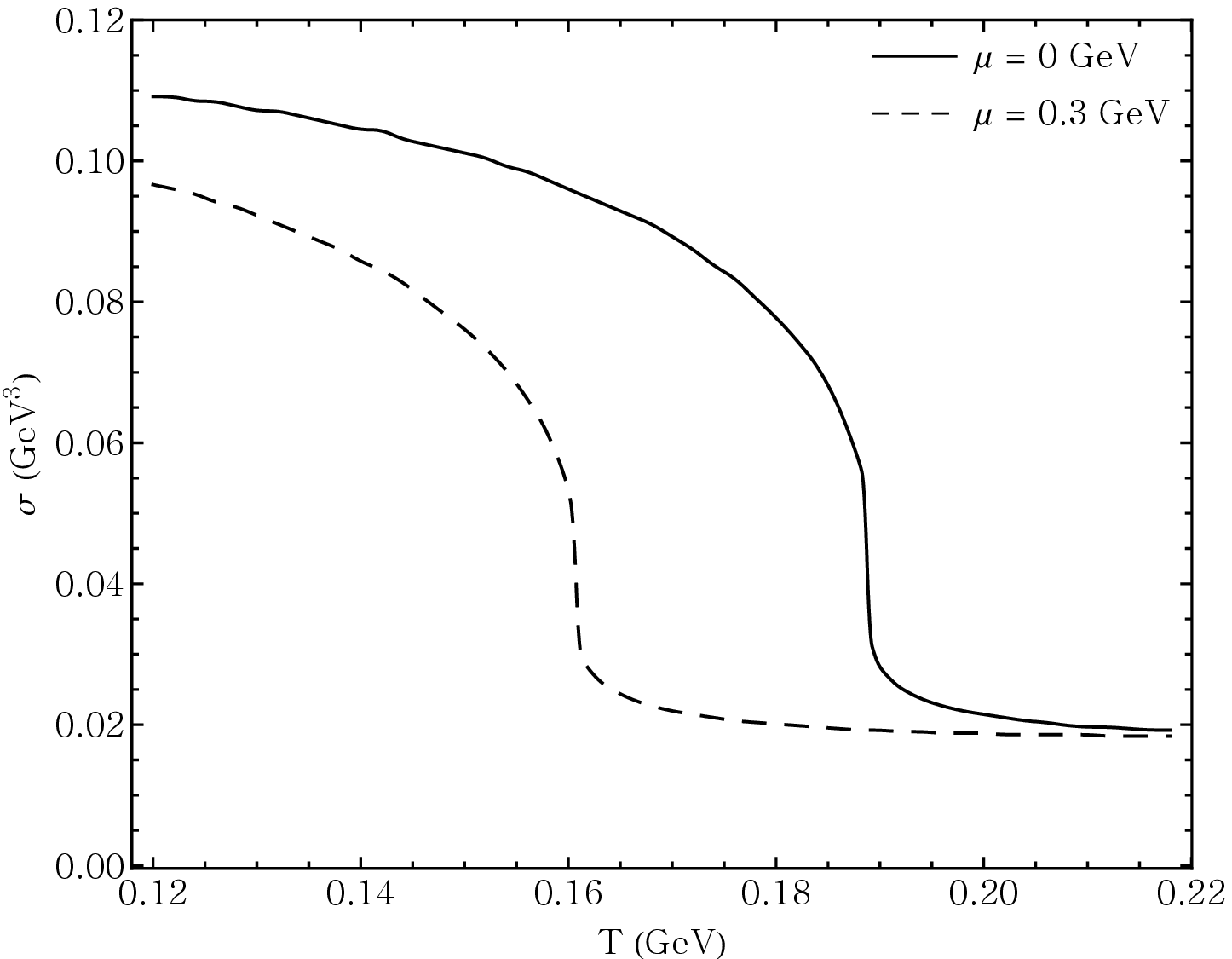}
}
\subfloat[]{
  \includegraphics[width=0.33\textwidth]{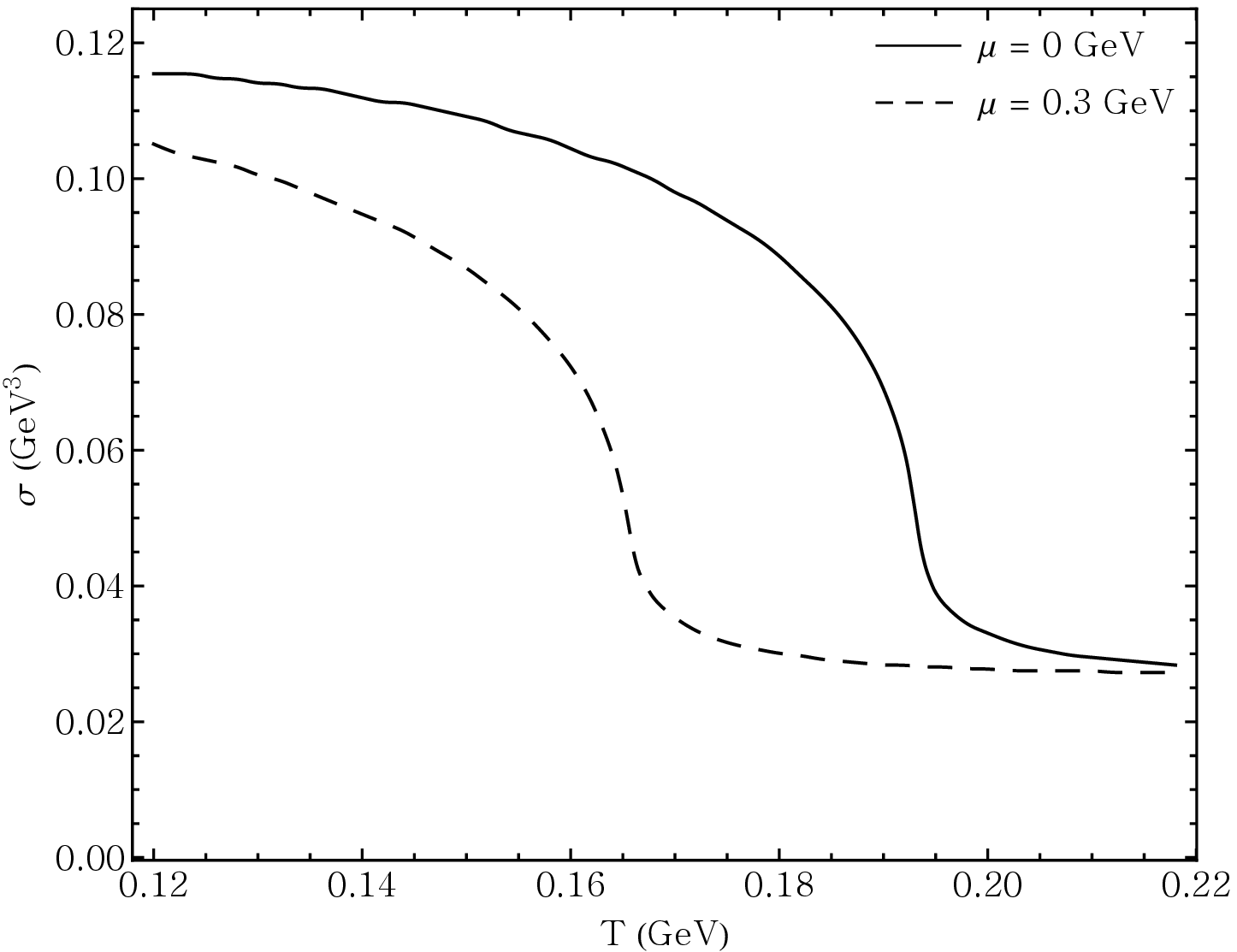}
}
\caption{Dependence of $\sigma$ on $T$, for zero and non-zero values of $\mu$ and (a) $m_l=m_s=10$ MeV, (b) $m_l=m_s=35$ MeV, and (c) $m_l=m_s=45$ MeV. In all cases, chemical potential has no effect on the order of the phase transition.} 
\label{symmetric}
\end{figure}

The critical value of the quark mass separating first-order and crossover phase transitions is  $m_q = 35$ MeV. Figure \ref{symmetric} shows examples of first-order, second-order, and crossover transitions for representative values of the quark mass and chemical potential. 
This critical quark mass is found to be independent of $\mu$, implying that this flavor-symmetric model cannot produce a critical point in the $T-\mu$ plane. 

%We also investigate the behavior of the transition at large values of the quark mass. In particular, we find that the model is limited to quark masses less than $150$ MeV, and cannot be used to analyze the upper right corner of the Colombia Plot. Figure \ref{puregaugeish} shows that when $m_q=100$ MeV, the transition is identifiable and clearly crossover. However, explicit symmetry breaking is dominant at high temperatures, where the values of $\sigma$ are bounded below by $0.1$ GeV$^3$. We find that for $m_q >> 100$ MeV, the chiral condensate increases monotonically as a function of temperature. An AdS/Glue model that incorporates realistic chiral symmetry breaking may be useful in examining the pure gauge limit where the transition is first order. 
%\begin{figure}[h!] 
%\centering
%\subfloat[]{
%  \includegraphics[width=0.5\textwidth]{puregauge.eps}
%}
%\caption{Dependence of $\sigma$ on $T$, for $\mu=0$ and large ($\geq 100$ MeV) values of the quark mass. Here $\sigma$ shows increasing behavior with no first order transition.} 
%\label{puregaugeish}
%\end{figure}

%%%%%%%%%%%%%%%%%%%%%%%%%%%%%%%%%%%%%%%%%%%%%%%%%%%%%%%
%%%%%%%%%%%%%%%%%%%%%%%%%%%%%%%%%%%%%%%%%%%%%%%%%%%%%%%

\subsection{2+1-flavors \label{2p1Flav}}

\begin{figure}[h!] 
\centering
\subfloat[]{
  \includegraphics[width=0.45\textwidth]{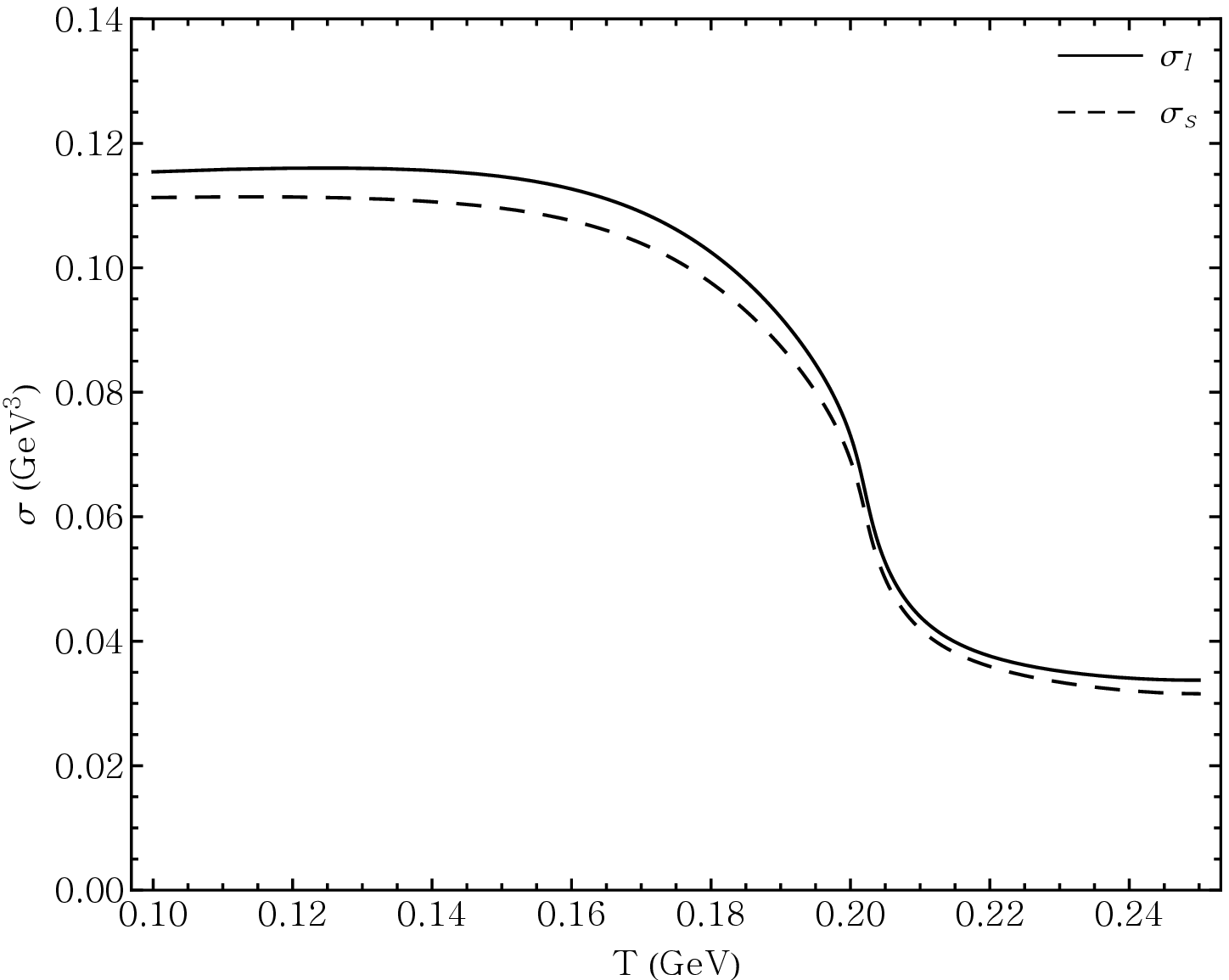}
}
\subfloat[]{
  \includegraphics[width=0.45\textwidth]{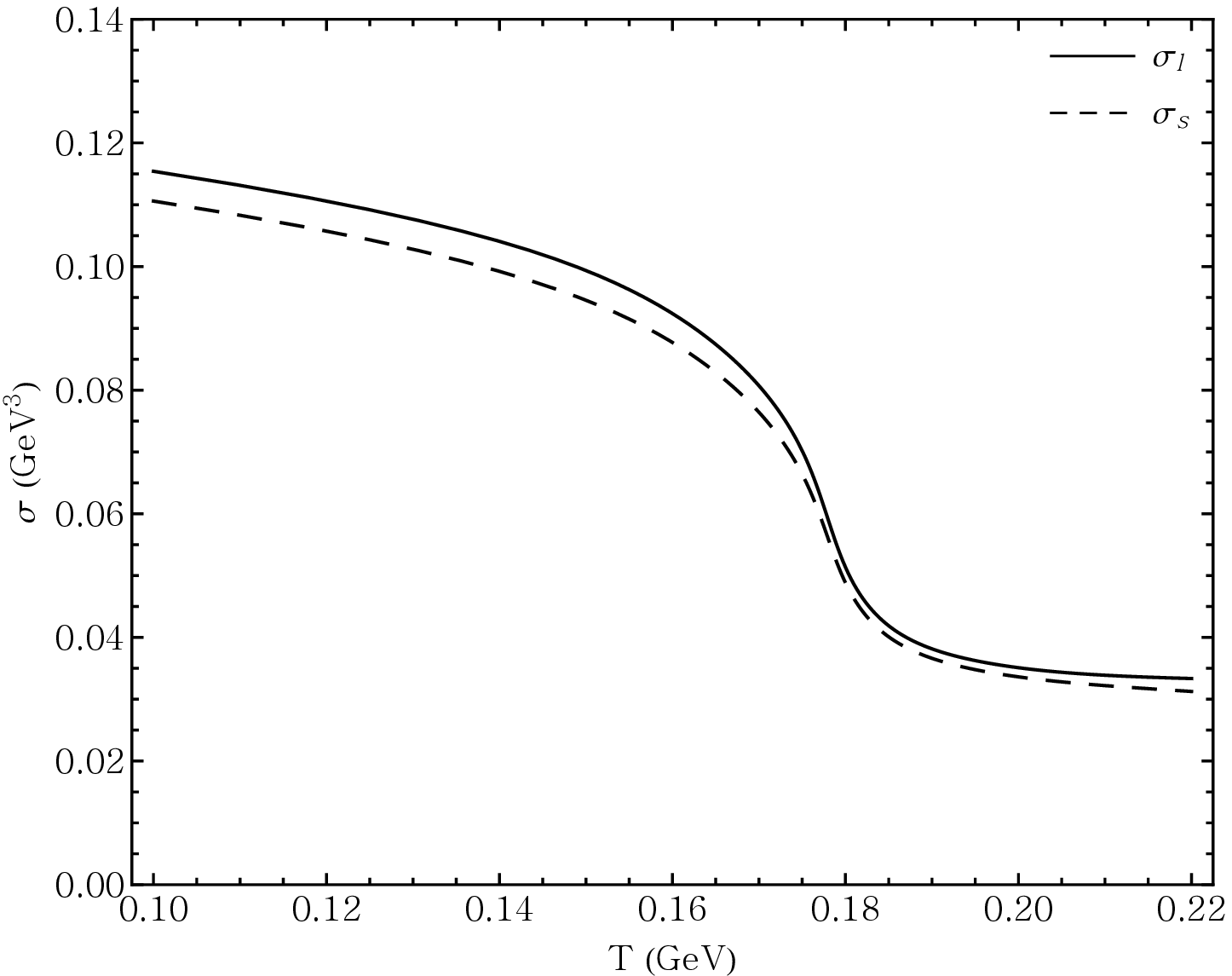}
}
\caption{Dependence of $\sigma_l$ and $\sigma_s$ on $T$ with $m_l=40$ MeV, $m_s=70$ MeV and (a) $\mu=0$ MeV and (b) $\mu=300$ MeV.} 
\label{asym1}
\end{figure}

Finding no critical point in the flavor-symmetric case, we extend the model to 2+1 flavors. 
The shooting is extended to find values of $\sigma_l$ and $\sigma_s$ that yield regular solutions for both chiral fields. Figure \ref{asym1} shows a representative case for $m_l=40$ MeV, $m_s=70$ MeV at $\mu=0$ MeV and $\mu=300$ MeV. As in the flavor-symmetric case, the order of the phase transition is not affected by the chemical potential. 
%From Figure\ref{asym1} it is clear that $\sigma_l$ and $\sigma_s$ do not coincide. In fact, by examining the case where $m_l=0$ MeV, $m_s = 200$ MeV, we find that $\sigma_l$ undergoes a second-order transition at all values of $\mu$ whereas the phase transition for $\sigma_s$ is first-order. Thus we find that the order of the transition is not necessarily the same for the two flavors, though no combination of quark masses produces a first-order transition for one flavor and crossover for the other. In the case that the orders do not agree, we characterize the transition as second-order. 

%When $m_s$ becomes large, we find second order phase transitions as $m_l \rightarrow 0$, in agreement with the chiral limit of two flavor QCD, where the transition belongs to the $O(4)$ universality class \cite{PhysRevD.29.338}. CAN WE ACTUALLY MAKE THIS CLAIM?? 

\begin{figure}[htb] 
\centering
\subfloat[]{
  \includegraphics[width=0.5\textwidth]{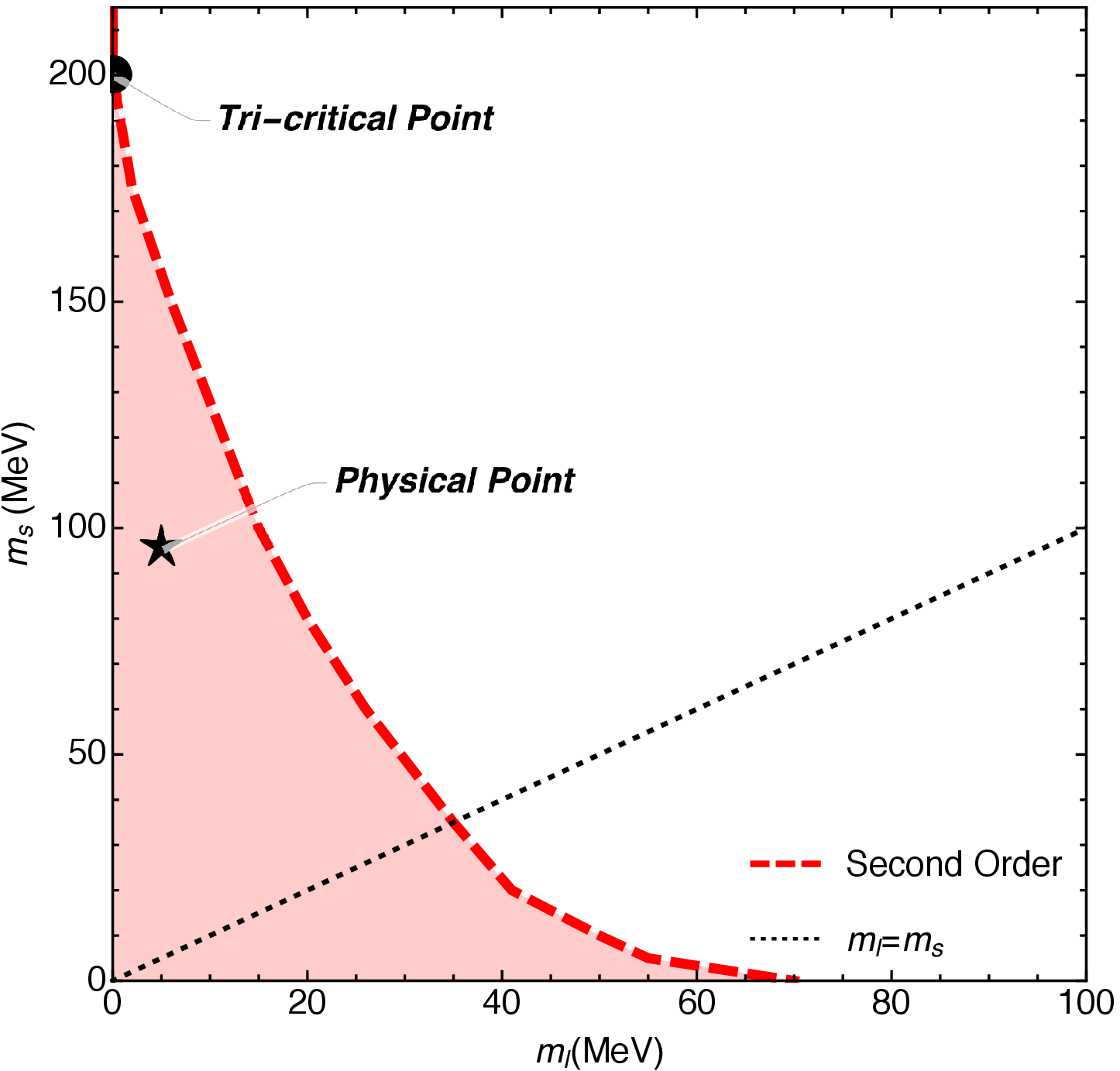}
}
\subfloat[]{
  \includegraphics[width=0.5\textwidth]{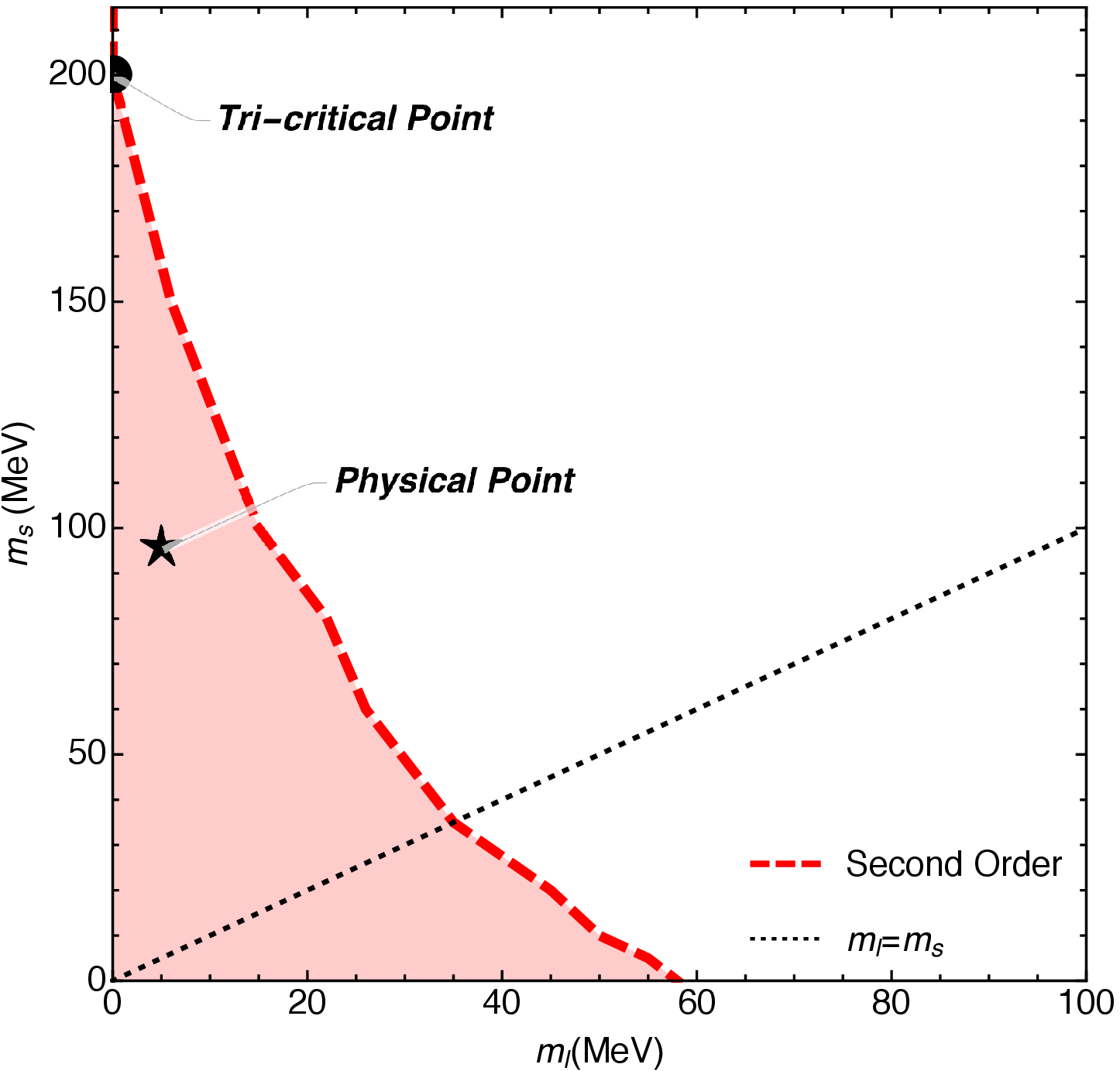}
}
\caption{The Columbia Plot for (a) $\mu=0$ MeV and (b) $\mu=300$ MeV. The shaded region indicates mass values where a first-order phase transition is produced, and the dashed line shows the line of second-order phase transitions separating it from the crossover region. The dotted line indicates the line of $m_s=m_l$.} 
\label{massplane}
\end{figure}

%\section{Columbia Plot}
The chiral phase structure at a particluar $\mu$ is summarized by finding the second-order line in the Columbia Plot. Figure \ref{massplane} shows representative Columbia Plots for $\mu=0$  and $\mu=300$ MeV. 
We find the tri-critical point, marking the end of the first-order region, at $m_s=200$ MeV. At greater $m_s$ values, second-order transitions occur at $m_l=0$  and rapid crossover transitions occur for finite values. 
The physical point $m_l = 5$ MeV, $m_s = 95$ MeV is within the first-order region, contradicting expectations from lattice results at $\mu = 0$. 
The location of the second-order line may be adjusted by the choice of input parameters, but this work is focused on the dependence of the phase transition order on chemical potential.

At small $m_s$ there are deviations in second-order line the on the order of 1 MeV as $\mu$ is varied. However, these small changes are likely numerical artifacts rather than evidence of a qualitative change.
Thus, the phase transition order is unaffected by $\mu$, indicating that the critical surface depicted in Figure \ref{critical_surface} has zero curvature in this model, and no critical point in the $T-\mu$ plane is possible. 

\subsection{Chiral potential}\label{chiPotential}
\begin{figure}[htb] 
\centering
\subfloat[]{
\includegraphics[width=0.6\textwidth]{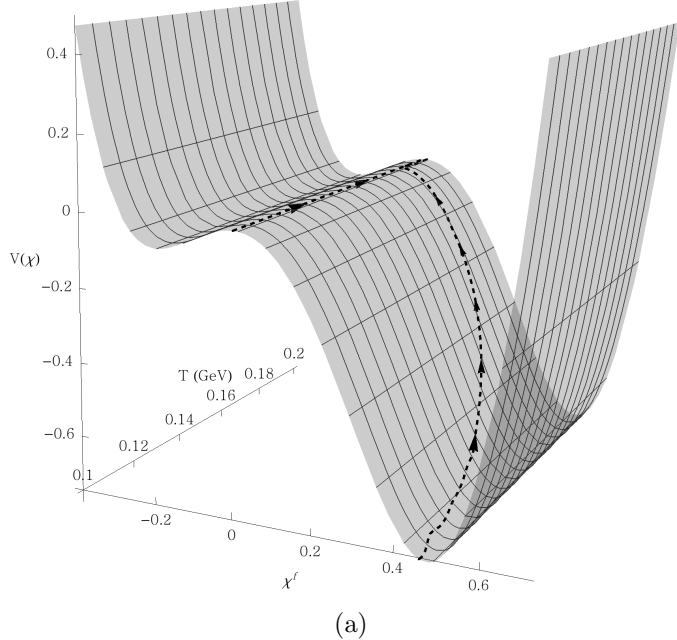}
}
\caption{Projection of $\chi^f$ onto $V(\chi)$ as a function of temperature for the case $m_l=m_s = 10$ MeV and $\mu = 0.010$ GeV. Symmetry restoration is clear at $T_c \sim 0.185$ GeV. } 
\label{symmetryrestoration1}
\end{figure}

Further analysis reveals how the solutions to (\ref{symmetricEOM}) relate to the flavor-symmetric scalar potential (\ref{chipotential}). In particular, the effects of spontaneous symmetry breaking is evident from the vacuum states of potential. We examine $V(\chi^f)$, where $\chi^f \equiv  \chi(u\rightarrow 1)$ is the near-horizon value of the chiral field. Figure \ref{symmetryrestoration1} shows the temperature-independent potential for a representative case with $m_l=m_s=10$ MeV and $\mu=10$ MeV. The near-horizon solution is projected onto the potential, showing first-order chiral symmetry restoration as $T \rightarrow T_c$. The trivial solution $\chi(u)=0$ is also plotted, but below $T=T_c$ it is  energetically disfavored.
Because $m_q$ is small, the effects of explicit symmetry breaking are small, and at $\chi^f = 0$, chiral symmetry is restored. 
The potential is  asymmetric in $\chi^f$, in contrast to the 2-flavor model which also permits negative solutions for $\chi,\, \sigma$. 

%At low temperatures the near-horizon value of the chiral field minimizes $V(\chi)$. As temperature increases, $\chi^f$ no longer inhabits the minimum of the potential and approaches the zero solution. 

\begin{figure}[h!] 
\centering
  \includegraphics[width=0.6\textwidth]{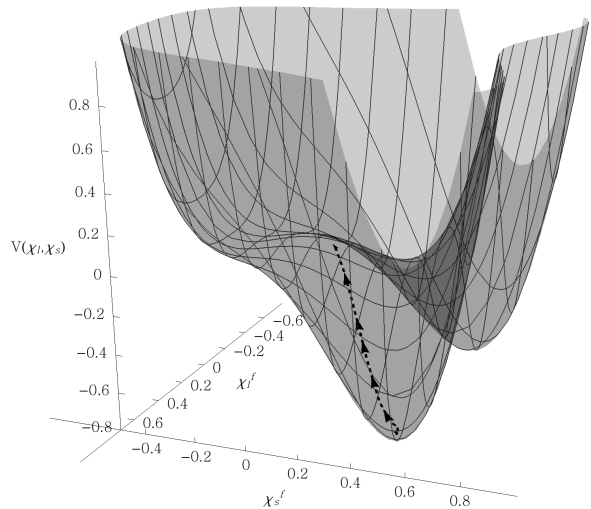}
\caption{Projection of $\chi_l^f$ and $\chi_s^f$ onto the surface of $V(\chi_l,\chi_s)$, parameterized by temperature in the range $T = [0.1,0.23]$ GeV.} 
\label{symmetryrestoration2}
\end{figure}

%\subsection{Analysis of the Potential}
We perform a similar analysis for the 2+1-flavor potential (\ref{chipotential}). With three flavors, we now project the temperature dependence of $\chi_l^f$ and $\chi_s^f$ onto the surface given by $V(\chi_l,\chi_s)$. Figure \ref{symmetryrestoration2} shows a representative case with $m_l=30$ MeV, $m_s=100$ MeV and $\mu=0.1$ GeV. The path along the surface is parameterized by temperature, ranging from $0.1$ GeV to $0.23$ GeV. The transition is crossover and the chiral fields do not vanish, a result of explicit chiral symmetry breaking due to the nonzero quark masses. As in the two-flavor case, when $T \rightarrow T_c$ the effects of spontaneous symmetry breaking disappear. 

\section{Conclusion}
In this work, we investigate the chiral phase transition in 2+1-flavor soft-wall AdS/QCD at finite temperature and quark chemical potential.
The scalar VEV is modified to include strange quarks, and higher order terms in the scalar potential are included in the action. A quartic term allows for independent sources of explicit and spontaneous chiral symmetry breaking, while a cubic t'Hooft determinant term allows for flavor mixing and first-order phase transitions. 
All analysis is performed in the finite chemical potential regime, using the AdS--Reissner-Nordstr{\"o}m metric.
Using the shooting method we numerically solve for the chiral field and extract the dependence of the chiral condensates on temperature and chemical potential. 

In the flavor-symmetric case where $m_l=m_s$, we find  critical quark mass $m_c = 35$ MeV separates first-order from crossover transitions. 
In agreement with lattice results and other nonperturbative methods, we find the second-order curve that separates first-order from crossover transitions for the flavor asymmetric case. When $m_s$ is above the tri-critical point, the results are consistent with the two flavor model, where the transition is second-order at $m_l=0$ and crossover otherwise. 

This paper presents improvements upon earlier results by incorporating finite chemical potential in a  2+1-flavor model, enabling exploration of the full chiral dynamics. 
Because the this holographic model admits no critical point in the $T-\mu$ plane. 
%Finite baryon number density has no effect on the order of the transition, regardless of quark mass.
%In particular this model couples both quark flavors to the chemical potential, and is qualitatively symmetric between $T$ and $\mu$. 
Future work in this area should provide qualitative differences at finite quark chemical potential. One possibility is to couple quark chemical potential to the light quarks only. A modified black hole metric may also introduce qualitative differences between the effects of temperature and chemical potential.
Finally, for a self-consistent model, the dilaton and black hole metric should be solved dynamically from the gravity action, rather than being parameterized by hand.

%\section*{Acknowledgments}
%The authors would like to thank Macalester College for funding this work.

%%%%%%%%%%%%%%%%%%%%%%%%%%%%%%%%%%%

\bibliographystyle{utphys.bst}
\bibliography{SlashTheoryEprint1.bib}

% ø\_(:-])_/ø  ø\_(:-])_/ø
% A Small Purple Boat Production
\end{document}